\documentclass[epj,referee]{svjour}

\begin{document}

\title{Non-equilibrium spin accumulation in ferromagnetic single-electron transistors}
\author{Arne Brataas\inst{1,2} \and Yu. V. Nazarov\inst{1} \and J. Inoue\inst{1,3}
\and Gerrit E.W. Bauer\inst{1}}
\institute{Delft University of Technology, Laboratory of Applied Physics and 
Delft Institute of Microelectronics and Submicrontechnology (DIMES),
2628 CJ Delft, The Netherlands \and
Philips Research Laboratories, Prof. Holstlaan 4, 5656 AA Eindhoven, The
Netherlands
\and Nagoya University, 
Department of Applied Physics, Nagoya,
Aichi 46401, Japan}
\offprints{Arne Brataas}
\mail{arneb@duttnto.tn.tudelft.nl}

\abstract{
We study transport in ferromagnetic single-electron transistors. 
The non-equilibrium spin accumulation on the island caused by a finite
current through the system is described by a generalized theory of the
Coulomb blockade. It enhances the tunnel magnetoresistance and has a 
drastic effect on the time-dependent 
transport properties. A transient decay of the spin accumulation may 
reverse  the electric current on time scales of the order of the
spin-flip relaxation time.
This can be used as an experimental signature of the non-equilibrium spin 
accumulation.
}

\maketitle

\section{Introduction}

Single-electron tunneling has been an active area of research during
the last decade (for a review see Ref. \cite{Grabert92}). In a double
tunnel junction system, the Coulomb blockade effect is pronounced when
the charging energy of the island is larger than the thermal energy,
$e^{2}/2C>k_{B}T$ ($C$ is the capacitance of the island), and the
tunnel resistances are larger than the quantum resistance
$R>R_{K}=h/e^{2}$. So far most of the research on the double tunnel
systems has been in systems using metals (normal or superconducting)
or semiconductor hetero-structures \cite{Grabert92,Kastner92:849}. The
so-called orthodox theory of single-electron tunneling
\cite{Grabert92} has been very successful in explaining experiments
where the single-particle energy separation is smaller than the
thermal energy.

Quite unrelated, the giant magnetoresistance in magnetic
multilayers and the spin-tunneling magnetoresistance in
ferromagnet-insulator-ferromagnet systems have attracted great interest \cite
{Gijs97:285,Levy94:367}. In a magnetic multilayer or in a trilayer "spin-valve"
structure the magnetization of the adjacent magnetic layers may vary. 
When the layers are antiparallel, an external magnetic field can align 
the magnetization of the ferromagnetic layers. The observed decrease in 
the resistance is a result of  
spin-dependent scattering in the materials.

In recent experiments Ono et al. studied the properties of ferromagnetic
single-electron transistors (FSETs) \cite{Ono96:3449,Ono97:1261}. This is 
a new and interesting system in which both the charging aspect
due to the Coulomb energy of the excess charge on the island and the
magnetoresistance due to the spin-dependent tunneling rates are of
 importance. Ono et al. found an enhanced magnetoresistance in
the Coulomb Blockade regime and a monotonic phase shift of the
Coulomb oscillations induced by the magnetic field. The enhancement of the
magnetic valve effect in the Coulomb blockade regime has been ascribed to
co-tunneling, which is of a higher order in the tunneling resistances and so
the difference in the spin-dependent tunneling between the 
antiparallel and
the parallel configuration is larger \cite{Ono97:1261,Takahashi98:1758}. The
magneto-Coulomb oscillations 
can be explained in terms of changes in the free energy of the island 
electrode and the leads in the presence of an external magnetic field \cite
{Shimada97:11075}. Coulomb charging effects have also been seen in
discontinuous multilayers \cite{Sankar97:5512} and in small cobalt clusters 
\cite{Schelp97:R5747}.

A generalization of the orthodox theory to describe a FSET by
introducing spin-dependent tunneling rates has been presented by Barnas and
Fert \cite{Barnas98:1058} and Majumdar and Hershfield \cite{Majumdar97:09206}.
However they neglected the effects of a non-equilibrium spin accumulation on
the island caused by the spin-dependent tunneling rates, which 
can have a drastic effect on the transport properties of the ferromagnetic
single-electron transistor.

In a ferromagnetic single-electron transistor, the transition rates
for tunneling into or out of the island depend on the electron spin.
An electric current through the system therefore implies a spin
current out of or into the island $(\partial s/\partial
t)_{\mbox{tr}}$. This creates a non-equilibrium excess spin on the
island. The excess spin $s$ decays by spin-flip relaxation so that
$s=(\partial s/\partial t)_{\mbox{tr}}\tau_{\mbox{sf}}$, where $\tau
_{\mbox{sf}}$ is the spin-flip relaxation time.  The non-equilibrium
spin accumulation on the island is equivalent to a non-equilibrium
chemical potential difference $\Delta \mu $ between the spin-up and
the spin-down states. In the case of a normal metal island this
chemical potential difference can be evaluated as $\Delta \mu =s\delta
$, where $\delta $ is the single-particle energy spacing at the Fermi
energy (inverse density of states). The chemical potential difference
modifies the transport properties.  The non-equilibrium spin
accumulation is important for the transport properties of the FSET
when the non-equilibrium chemical potential difference $\Delta \mu $
is of the same order as the Coulomb charging energy
$E_{c}=e^{2}/2C$. The spin current is of the same order as the
electric current, $e(\partial s/\partial t)_{\mbox{tr}}\sim I$. We are
interested in voltages of the order of the Coulomb gap, so we have
$I\sim GE_{c}/e$, where $G$ is the typical junction conductance. From
these crude considerations we deduce that the non-equilibrium spin
accumulation is important when the spin-flip relaxation time and/or
the single-particle energy separation are sufficiently large,
i.e. when
\begin{equation}
\tau _{\mbox{sf}}\delta /h>R/R_{K},  \label{condition}
\end{equation}
irrespective of $E_c$.

The spin-flip relaxation time is crucial for the observation of the
non-equilibrium spin accumulation. The spin-flip scattering in clean
metals at low temperatures is dominated by the spin-orbit scattering,
$\tau_{\mbox{sf}}\approx \tau _{\mbox{so}}$. Since spin-orbit coupling
is a relativisitic effect, it increases with the atomic number $Z$ of
the metal.  The single-particle energy spacing on the island increases
with decreasing number of atoms $N$ on the island as $\delta \sim
E_{F}/N$, where $E_{F} $ is the Fermi energy. Thus the size of the
island must be small for heavy elements in order to observe big
effects of the spin-accumulation. Let us first estimate the spin-flip
relaxation time in an island with an ideal clean surface, where the
spin-flip scattering at the boundary of the island can be
disregarded. The spin-flip relaxation time in single-crystal aluminium
is ($\tau_{\mbox{sf}} \sim 10^{-8}s$ at $T=4.3K$ \cite{Johnson85:1790}
( $\tau _{\mbox{sf}}\sim 10^{-10}s$ in polycrystalline aluminium
\cite{Tedrow94:174}) and $\tau _{\mbox{sf}}\sim 10^{-11}s$ for
gold\cite{Johnson93:2142}. The Fermi energy is $E_{F}\sim 10eV$, and
the spin accumulation may therefore be expected to play a significant
role in an Al island with less than 10$^{8}$ atoms (10$^{6}$ atoms in
polycrsalline aluminium).  In the limit $\tau _{\mbox{sf}}\rightarrow
0$ or for a large island, our results reduce to the models in
Refs. \cite{Barnas98:1058,Majumdar97:09206}, where the spin was
assumed to be equilibrated.

Spin-accumulation is important in small systems also when the boundary
roughness gives a dominant contribution to the spin-flip
scattering. In such a case, the spin-orbit relaxation time is $\tau
_{\mbox{so}}=f\tau ,$ where $f$ is the ratio of non spin-flip to
spin-flip scattering rates and $\tau $ is the momentum relaxation
time. For a small system with an ideally rough surface $\tau \sim
L/v_{F},$ where $L$ is the size of the system and $v_{F}$ is the Fermi
velocity. Abrikosov and Gor'kov calculated \cite{Abrikosov62:1088}
$f=(\alpha Z)^{-4}$, where $\alpha =e^{2}/(\hbar c)\simeq 1/137$ is
the fine-strucure constant, in rough agreement with experiments \cite
{Meservey78:805}. Using $L\sim aN^{1/3}$, where $a\sim 1/k_{F}$ is the
lattice constant $\tau \sim N^{1/3}/(\hbar E_{F})$ and the relation
(\ref {condition}) reads for small particles with rough boundaries:
\begin{equation}
N<\left( f\frac{R_{K}}{R}\right) ^{3/2}.
\end{equation}
Assuming that the tunnel resistance is of the order of the quantum
resistance (e.g. in the measurement by Ono et al. \cite{Ono97:1261}
$R/R_{K}\sim 1$), and using estimates for the ratio of non spin-flip
to spin-flip scattering rates from the measurements in Ref. \cite
{Meservey78:805}, we expect to see effects of a non-equilibrium spin
accumulation in Cu ($\log f\sim 2.1-2.4$) when $N<$ $10^{3}$, in Al
($\log f\sim 2.8-3.6$) when $N<\allowbreak 10^{4}-10^{5}$, in Na
($\log f\sim 4.2-5.3$) when $N<\allowbreak 10^{6}-10^{9}$ and in Li
($\log f\sim 4.9-7$) when $N<\allowbreak 10^{7}-10^{10}$. For smooth
boundaries, these numbers should be correspondingly larger. For our
purposes, the study of effects of a non-equilibrium spin accumulation,
one should choose a light metal with a long spin-orbit relaxation time
(a large parameter $f$).

Note that the orthodox model is still valid for the small clusters
described above. For a small island, the capacitance scales like
$C\sim aN^{1/3}$. The ratio between the single-particle energy spacing
and the Coulomb charging energy is therefore $\delta /E_{C}\sim \left(
E_{F}a/e^{2}\right) N^{-2/3}$.  The prefactor is $\left(
E_{F}a/e^{2}\right) \sim 1$, and thus the single-particle spacing is
much smaller than the Coulomb charging energy.  Our generalized
orthodox model is therefore valid when the temperature is larger than
the single-particle energy spacing. For lower temperatures, the
single-particle energy spacing will also appear in the current-voltage
characteristics.

''Modern'' metals, like arm-chair nanotubes \cite{Tans98:49} or (magnetic)
semiconductor heterostructures \cite{Matsukara98:R2037} can also be
interesting as island materials. The first because of a possible huge
spin-flip relaxation time ($Z=6$ for carbon) and the latter since islands
containing a small number of electrons can be created by depletion of the
two-dimensional electron gas \cite{Grabert92}. However, in these systems
quantum size effects start to play a role which have our attention.

In small tunneling systems where Eq. (\ref{condition}) is satisfied, the spin-flip
relaxation time is longer than the charge relaxation time $RC$. This can be
seen from Eq. (\ref{condition}), $\tau _{\mbox{sf}}>(2E_{c}/\delta
)RC $, and noting that the charging energy is larger than the
single-particle energy spacing for all but in few-electron systems. Hence the
long-time response of the system is dominated by the spin dynamics. We will
demonstrate that this feature can be used as an experimental signature 
of the non-equilibrium spin accumulation.

The paper is organized in the following way. In the next section \ref{s:gen}   
 we model the transport in the ferromagnetic single-electron transistor by
generalizing the orthodox theory. The simple case of a halfmetallic
ferromagnet-normal metal-halfmetallic Ferromagnetic (HF/N/HF) FSET can be
treated analytically and is considered in section \ref{s:half}. The
transport properties in the general case are investigated numerically in
section \ref{s:num}. Finally our conclusions are presented in section \ref
{s:con}. Selected results of the present work have been presented in
Ref. \cite{Brataas98}.

\section{Model}

\label{s:gen}We consider the situation shown in Fig.\ \ref{f:set}. A
(ferromagnetic or normal) metal island is attached to two 
(ferromagnetic or normal) leads by two
tunnel junctions. There is an applied source-drain voltage $V$ between the
right and the left reservoir and a gate voltage source coupled capacitively
to the island. We assume collinear magnetization in the leads and the
island. The direction of the magnetizations in the island and the right lead
can be parallel or antiparallel to the direction of the magnetization in the
left reservoir as depicted by the broken arrows in the Fig.\ (\ref{f:set}).
When the coercive fields of the magnets are different 
different configurations of the magnetizations  
can be realized by sweeping an external magnetic field.
 We ignore for simplicity the complications due to the
possible appearance of a magnetic domain structure in the ferromagnets.
The tunnel junctions have a
capacitance $C_{i}$ and spin-dependent conductances $G_{i\sigma },$ where $i=1,2$ denotes the first and the second junction and $\sigma $ denotes up $(+)$ or down $(-)$ spin electrons. The spin-dependent conductance of the
first junction, $G_{1\sigma }$, is proportional to the spin-dependent
density of states in the left reservoir at the Fermi level $N_{\mbox{L}\sigma }(0)$, the island at the
Fermi level $N_{\mbox{I}\sigma }(0)$ and the angular averaged spin-dependent
tunneling probability at the Fermi level, $G_{1\sigma }=2\pi e^{2}N_{L\sigma
}(0)N_{I\sigma }(0)T_{1\sigma }(0)/\hbar $ \cite{Julliere75:225}.
 In a ferromagnet the tunneling
probability is significantly different for the $s$- and $d$-electrons. 
The density of states and the tunneling probabilities  
should therefore be regarded as phenomenological parameters 
which are to be constant 
in an interval of the order of the
applied source-drain voltage at the Fermi surface. This 
assumption has been found to be valid if the 
applied source-drain voltage is lower than 100 mV \cite{Barnas98:1058}. 
The expression for the
spin-dependent conductance of the second junction is similar. We take 
the gate capacitance $C_{g}$ to be small compared to the junction
capacitances $C_{1}$ and $C_{2}$ and other impedances are disregarded.
The energy difference associated with the tunneling of one
electron into the island through junction $i$ is \cite{Grabert92} 
\[
E_{i}(V,q)=\kappa _{i}eV+\frac{e(q-e/2)}{C_{1}+C_{2}}, 
\]
where $q=-ne+q_{0}$ is the charge of the island, the total capacitance
is $ 1/C=1/C_{1}+1/C_{2}$ and $\kappa _{i}=C/C_{i}$. The number of
excess electrons on the island is $n$. The offset charge $q_{0}$ is
controlled by the gate voltage coupled to the island by the gate
capacitance, $q_{0}=C_{g}V_{g}$.

At a finite current through the island, the energy diagram 
in Fig.\ \ref{f:non} should be considered. Here we show the equilibrium
chemical potentials in the left and the right reservoir and the
non-equilibrium chemical potentials for the spin-up and the spin-down
electrons in the island. It is assumed that the energy relaxation in the
island is much faster than the time between the tunneling events $\sim RC$,
so that the distributions of the energy levels for the different spins are
described by Fermi functions. The non-equilibrium chemical potentials in the
island are spin-split by $\Delta \mu $ due to the spin accumulation. The
non-equilibrium chemical potential difference causes novel phenomena on
the single-electron transistor to be discussed below.  

The tunnel conductances are taken to be much smaller than the quantum
conductance $G_{i\sigma }\ll G_{K}=1/R_{K}$. Thus we disregard
co-tunneling \cite{Takahashi98:1758} and calculate the current-voltage
characteristics in the system by using the semiclassical master equation.
The transition rates can be found from Fermi's Golden Rule \cite{Grabert92}. The spin-dependent tunneling rates are 
\begin{eqnarray}
\overrightarrow{\Gamma ^{1\sigma }}_{n+1,n} &=&\frac{1}{e^{2}}G_{1\sigma
}F(E_{1}(V,q)-\sigma \Delta \mu /2),  \label{rates} \\
\overleftarrow{\Gamma ^{1\sigma }}_{n,n+1} &=&\frac{1}{e^{2}}G_{1\sigma
}F(-E_{1}(V,q)+\sigma \Delta \mu /2),  \nonumber \\
\overleftarrow{\Gamma ^{2\sigma }}_{n+1,n} &=&\frac{1}{e^{2}}G_{2\sigma
}F(E_{2}(-V,q)-\sigma \Delta \mu /2),  \nonumber \\
\overrightarrow{\Gamma ^{2\sigma }}_{n,n+1} &=&\frac{1}{e^{2}}G_{2\sigma
}F(-E_{2}(-V,q)+\sigma \Delta \mu /2).  \nonumber
\end{eqnarray}
$\overrightarrow{\Gamma ^{1\sigma }}_{n+1,n}$ denotes
transition from the left reservoir to the island, so that the number of
electrons on the island is changed from $n$ to $n+1$, etc.  and
\[
F(E)=\frac{E}{1-\exp (-E/k_{B}T)},
\]
where $k_{B}T$ is the thermal energy. We also define the total forward
rates $\overrightarrow{\Gamma ^{i}}=\overrightarrow{\Gamma ^{i\uparrow
}}+ \overrightarrow{\Gamma ^{i\downarrow }}$, where $i=1,2$, and
analogous backward rates. The combined rate for an increase of the
number of excess electrons on the island is $\Gamma
_{n+1,n}=\overrightarrow{\Gamma ^{1}}_{n+1,n}+\overleftarrow{\Gamma
^{2}}_{n+1,n}$ and analogous $ \Gamma _{n,n+1}=\overleftarrow{\Gamma
^{1}}_{n,n+1}+$ $\overrightarrow{ \Gamma ^{2}}_{n,n+1}$. The master
equation which determines the probability $ p_{n}$ of having $n$
excess electrons on the island is
\begin{eqnarray}
\frac{dp_{n}}{dt} &=&-p_{n}\left( \Gamma _{n-1,n}+\Gamma _{n+1,n}\right) 
\label{master} \nonumber \\
&&+p_{n+1}\Gamma _{n,n+1}+p_{n-1}\Gamma _{n,n-1}.
\end{eqnarray}
In the stationary state $dp_{n}/dt=0$ and the detailed balance
symmetry gives $\Gamma _{n+1,n}p_{n}=\Gamma _{n,n+1}p_{n+1}$. The current
through the first junction is $I_{1}=(I_{1}^{\uparrow }+I_{1}^{\downarrow })$
, where the current of electrons with spin $\sigma $ is 
\[
I_{1}^{\sigma }=e\sum_{n}p_{n}\left( \overrightarrow{\Gamma ^{1\sigma }}
_{n+1,n}-\overleftarrow{\Gamma ^{1\sigma }}_{n-1,n}\right) .
\]
In the second junction the current is $I_{2}=(I_{2}^{\uparrow
}+I_{2}^{\downarrow })$, 
\[
I_{2}^{\sigma }=e\sum_{n}p_{n}\left( \overrightarrow{\Gamma ^{2\sigma
}} _{n-1,n}-\overleftarrow{\Gamma ^{2\sigma }}_{n+1,n}\right) .
\]
The transport of spins into the island is determined by 
\begin{eqnarray}
\left( \frac{ds}{dt}\right) _{\mbox{tr}} &=&\left( \frac{d\left( N_{\uparrow
}-N_{\downarrow }\right) }{dt}\right) ^{\mbox{in}}-\left( \frac{d\left(
N_{\uparrow }-N_{\downarrow }\right) }{dt}\right) ^{\mbox{out}}
\label{spintr} \\
&=&(I_{1}^{\uparrow }-I_{2}^{\uparrow }-I_{1}^{\downarrow
}+I_{2}^{\downarrow })/e.  \nonumber
\end{eqnarray}
In the stationary situation we can use the current conservation $\
I_{1}^{\uparrow }+I_{1}^{\downarrow }=I_{2}^{\uparrow }+I_{2}^{\downarrow }$
and find 
\begin{equation}
\left( \frac{ds}{dt}\right) _{\mbox{tr}}^{\mbox{st.}}=2(I_{1}^{\uparrow
}-I_{2}^{\uparrow })/e=2(I_{2}^{\downarrow }-I_{1}^{\downarrow })/e.
\label{spinst}
\end{equation}
The spin balance is 
\begin{equation}
\frac{ds}{dt}=\left( \frac{ds}{dt}\right) _{\mbox{tr}}+\left( \frac{ds}{dt}
\right) _{\mbox{
rel}},  \label{spineq}
\end{equation}
where $(ds/dt)_{\mbox{ rel}}$ is the spin-flip relaxation rate. In
equilibrium there are $s_{0}$ spins on the island (for a normal metal
island $s_{0}=0$). The chemical potential is $\mu _{0}$. The
non-equilibrium spin-dependent chemical potentials are $\mu _{\uparrow
}=\mu _{0}+\delta \mu +\Delta \mu /2$ and $ \mu _{\downarrow }=\mu
_{0}+\delta \mu -\Delta \mu /2$. The total number of spins on the
island is $s=s_{0}+[(\delta \mu +\Delta \mu /2)\delta _{\uparrow
}^{-1}-(\delta \mu -\Delta \mu /2)\delta _{\downarrow }^{-1}]$, where
$\delta _{\sigma }^{-1}$ is the spin-dependent density of states. The
spin-flip relaxation rate is
\begin{equation}
\left( \frac{ds}{dt}\right) _{\mbox{rel}}=-\frac{s-\overline{s}}{\tau
_{ \mbox{sf}}}=-\frac{\Delta \mu }{\tau _{\mbox{sf}}\delta }.
\label{spinrel}
\end{equation}
In the limit of fast spin-flip relaxation ($\tau _{\mbox{sf}}\rightarrow
\infty $), the number of spins on the island is $\overline{s}=s_{0}+\delta
\mu (\delta _{\uparrow }^{-1}-\delta _{\downarrow }^{-1})$. Eq.\ (\ref
{spineq}) determines the non-equilibrium chemical potential shift $\Delta
\mu $. Here $\tau _{\mbox{sf}}$ is the spin-flip relaxation time in the
island and $\delta ^{-1}=(\delta _{\uparrow }^{-1}+\delta _{\downarrow
}^{-1})/2$ is the average density of states for spin up and spin down
electrons at the Fermi level in the island (the inverse single-particle
energy spacing). Eq. (\ref{spinrel}) effectively includes
many-body effects. The excess 
spin is related to the energy difference between spin-up and spin-down states by
$s-\overline{s}=\chi_s \Delta \mu / \mu_B^2$,
where $\chi_s$ is the spin susceptibility (for noninteracting electrons
$\chi_s=\mu_B^2\delta^{-1}$). In the stationary state ($dp_{n}/dt=0$, $ds/dt=0$), the
spin balance (\ref{spineq}) can be written as 
\begin{equation}
I_{s}=e\left( \frac{ds}{dt}\right) _{\mbox{tr}}=G_{s}2\Delta \mu /e,
\label{spmot}
\end{equation}
where the spin-flip conductance is defined by 
\begin{equation}
G_{s}=\frac{e^{2}}{2\tau _{s}\delta }.  \label{scond}
\end{equation}
From the equations for the tunneling rates (\ref{rates}) and the 
spin balance (\ref{spineq}), it is not directly obvious that there
is only one unique solution for the non-equilibrium chemical potential
difference $\Delta \mu $ for a given set of parameters (temperature, gate
voltage, source-drain voltage, capacitances and conductances). If Eq.\ (\ref
{spineq}) would be fulfilled for multiple solutions $\Delta \mu $ for a given set
of parameters, the current-voltage characteristics has a
hysteretic behavior, with a solution $\Delta \mu $
depending on the history. However, 
by extensive analytical and numerical studies with many different
parameters to be presented below we always found only one unique
solution of Eq.\ (\ref{spineq}). 

In this orthodox model the problem can be mapped on the equivalent circuit
in Fig.\ \ref{f:cir} by introducing a ''spin-flip capacitance'' 
$C_{s}\equiv e^{2}/2\delta $, so that 
\[
(es)/2=C_{s}(\Delta \mu /e),\mbox{ }\Delta \mu /s=e^{2}/(2C_{s})=\delta . 
\]
This ''charging energy'' of the spin-flip capacitance is thus simply the
single-particle energy cost of a spin-flip, $\delta $, or more generally,
the inverse of the magnetic susceptibility $\mu _{B}^{2}/\chi _{s}$.

Let us first discuss the situations in which $\Delta \mu$ vanishes
identically in the stationary
state. From the expressions for the spin transport (\ref{spintr}), the
current conservation through the two junctions, $I=I_{1}=I_{2},$ and the
spin-dependent rates (\ref{rates}), we see that when 
$\Delta \mu =0$ the spin-dependent
currents are related to the total current by 
\begin{eqnarray*}
I_{i}^{\uparrow } &=&\frac{1}{1+G_{i\downarrow }/G_{i\uparrow }}I, \\
I_{i}^{\downarrow } &=&\frac{1}{1+G_{i\uparrow }/G_{i\downarrow }}I,
\end{eqnarray*}
where $i=1,2$,
\[
I_s=2I\left( \frac{1}{1+G_{1\downarrow }/G_{1\uparrow
}}-\frac{1}{1+G_{2\downarrow }/G_{2\uparrow } }\right) .
\]
From this relation we can make the following observations. First, we
see that in the symmetric case, $G_{1\uparrow }/G_{1\downarrow
}=G_{2\uparrow }/G_{2\downarrow }$, the spin-current is zero for
$\Delta \mu =0$ and therefore $\Delta \mu =0$ is a solution of the
spin balance equation (\ref{spineq}) for any source-drain voltage $V$,
gate voltage $V_{g} $ and temperature $T$, i.e. \emph{there is a
solution with zero non-equilibrium chemical potential shift}. This
symmetry is expected for a structure of the type (Ferromagnet A-
Ferromagnet B - Ferromagnet A) with magnetization configuration
$\uparrow \uparrow \uparrow $ or $\uparrow \downarrow \uparrow $,
which is the case for the device measured in Refs. \cite
{Ono96:3449,Ono97:1261}. Thus in this case our theory reduces to those
in Refs. \cite{Barnas98:1058,Majumdar97:09206}, where the spins were
assumed to be equilibrated. Second, we find that for a general system
in the Coulomb blockade regime the current is zero, $I=0$, and hence a
vanishing non-equilibrium chemical potential difference $\Delta \mu
=0$ is a solution in the Coulomb blockade regime. This is physically
reasonable since it means that there is no spin accumulation on the
island when there is no current flowing through the junctions. The
Coulomb gap in the current-voltage characteristics is not modified by
the spin accumulation.

\section{Analytical results for a halfmetallic ferromagnet/normal metal/
halfmetallic ferromagnet FSET}

\label{s:half}We will now discuss an idealized case for which 
an analytic expression for the current-voltage can be derived at zero temperature. 
Let us assume that the leads are halfmetallic
ferromagnets and the island is a normal metal. The density of states at the
Fermi energy vanishes for the minority spins in a halfmetallic ferromagnet.
The capacitances are taken to be symmetric, $C_{1}=C_{2}=C$, and the gate 
voltage is
tuned so that the offset charge is zero $q_{0}=0$.
When the direction of the magnetizations in the ferromagnetic leads are
antiparallel the tunnel conductances are  
\begin{equation}
G_{1\uparrow }^{\mbox{AP}}=G_{1},G_{1\downarrow }^{\mbox{AP}}=0,G_{2\uparrow
}^{\mbox{AP}}=0,G_{2\downarrow }^{\mbox{AP}}=G_{2}.  \nonumber
\end{equation}
The conductances $G_{1\downarrow }^{\mbox{AP}}$ and $G_{2\uparrow
}^{\mbox{AP}}$ are zero, since there are no spin-down states at the
Fermi surface in the left reservoir and there are no spin-up states at
the Fermi surface in the right reservoir. The electric current is
$I=I_{1}^{\uparrow }$ and the spin-current is
\[
e\left( \frac{ds}{dt}\right) _{\mbox{tr}}=2(I_{1}^{\uparrow }-I_{2}^{\uparrow
})=2I. 
\]
The spin-current is directly
proportional to the current for any temperature $T$ and source-drain voltage 
$V$. In the absence of spin-flip
relaxation, $\tau _{\mbox{sf}}\rightarrow \infty $, the current through the
system vanishes, because an electron cannot propagate from the left
reservoir to the right reservoir without spin-flip.
The master equation can be solved exactly at zero temperature for specific 
voltages $V_{m}^{\mbox{AP}}$ \cite{Grabert91:143}:
\[
\frac{1}{2}\left( eV_{m}^{\mbox{AP}}-\Delta \mu \right) =E_{c}\left(
m+\frac{1}{2}\right),
\]
where $E_{c}=e^{2}/2C$ is the Coulomb charging energy and $m$ is an integral
number. The non-vanishing rates are 
\begin{eqnarray}
\overrightarrow{\Gamma ^{1\uparrow }}_{n+1,n} &=&\frac{1}{e^{2}}
G_{1}E_{c}(m-n)\Theta (m-n), \label{Rateferro} \\
\overleftarrow{\Gamma ^{1\uparrow }}_{n,n+1} &=&-\frac{1}{e^{2}}
G_{1}E_{c}(m-n)\Theta (n-m), \nonumber \\ \overleftarrow{\Gamma
^{2\downarrow }}_{n+1,n} &=&-\frac{1}{e^{2}} G_{2}E_{c}(m+n+1)\Theta
(-m-n-1), \nonumber \\ \overrightarrow{\Gamma ^{2\downarrow }}_{n,n+1}
&=&\frac{1}{e^{2}} G_{2}E_{c}(m+n+1)\Theta (m+n+1), \nonumber
\end{eqnarray}
where $\Theta (x)$ is the Heaviside function. The probability of finding $n$
electrons on the island is ($n\leq m$) \cite{Grabert91:143} 
\[
p_{n}(m)=\left( \frac{G_{1}}{G_{2}}\right) ^{n}\frac{(m!)^{2}}{
(m-|n|)!(m+|n|)!}p_{0}(m) 
\]
and $p_{0}(m)$ can be found from the normalization condition
$\sum_{n}p_{n}(m)=1$ \cite{Grabert91:143}. The current is
\[
I=G_{1,2}\left( V_{m}^{\mbox{AP}}-\frac{\Delta \mu +E_{c}}{e}\right) , 
\]
where we have defined the conductance for the two junctions in series,
$G_{1,2}=G_{1}G_{2}/(G_{1}+G_{2})$. The non-equilibrium chemical
potential difference is determined by Eq. (\ref{spmot}) which is
simplified to $I=I_{s}/2=G_{s}\Delta \mu /e$ as
\[
\Delta \mu =\frac{G_{1,2}}{G_{s}+G_{1,2}}\left(
eV_{m}^{\mbox{AP}}-E_{c}\right) .
\]
The current in the system can then be found to be 
\begin{equation}
I^{\mbox{AP}}=\frac{G_{s}G_{1,2}}{G_{s}+G_{1,2}}\left(
V_{m}-\frac{E_{c}}{e}\right) \label{APcur}
\end{equation}
at specific voltages determined by 
\[
eV_{m}^{\mbox{AP}}=E_{c}\left( 2m\frac{G_{s}+G_{1,2}}{G_{s}}+1\right) . 
\]
We see from the current (\ref{APcur}) that the Coulomb blockade
threshold is unchanged as expected. On entering the island from one of
the leads the electrons must flip their spins in order to be able to
tunnel through the other lead. The total conductance of the system is
thus given by the three resistances in series,
$1/G_{\mbox{tot}}^{\mbox{AP}}=1/G_{s}+1/G_{1}+1/G_{2} $ as can be seen
from the equivalent circuit in Fig. \ref{f:cir}. No instabilities
(i.e. multiple solutions of the non-equilibrium chemical potential
difference $\Delta \mu $ for a given set of parameters) exist.

For a parallel orientation of the magnetizations in the two leads,
where $\Delta \mu =0$, the current-voltage characteristics is \cite
{Grabert91:143} 
\begin{equation}
I^{\mbox{P}}=G_{1,2}(V_{m}^{\mbox{P}}-\frac{E_{c}}{e}),  \label{parcur}
\end{equation}
where $eV_{m}^{\mbox{P}}=E_{c}(2m+1)$. Note that the specific voltages for
which we have determined the current differ for the
parallel and the antiparallel situation. The junction magnetoresistance is
determined by 
\begin{equation}
\mbox{JMR}=(I^{\mbox{P}}-I^{\mbox{AP}})/I^{\mbox{P}}  \label{JMR}
\end{equation}
at the same voltages, which we cannot calculate exactly.
At high source-drain voltages at which the Coulomb charging effects can be
disregarded we see from Eq.\ (\ref{APcur}) and Eq.\ (\ref{parcur}) and Fig. 
\ref{f:cir} that the magnetoresistance for the system is 
\[
\mbox{JMR}=\frac{G_{1,2}}{G_{s}+G_{1,2}}. 
\]
In the limit of no spin-flip relaxation, $\tau _{\mbox{sf}}\rightarrow
\infty $ ($G_{s}\rightarrow 0$), the junction magnetoresistance is
100\thinspace \% since the current vanishes in the antiparallel
configuration and in the limit of perfect spin-flip relaxation, $\tau
_{\mbox{sf}}\rightarrow 0$ ($G_{s}\rightarrow \infty $), the junction
magnetoresistance vanishes.

\section{General configurations: Numerical results and discussions}

\label{s:num}In the general situation with arbitrary junction conductances
and capacitances and at finite temperatures, the spin-current and the electric
current in the system have to be calculated numerically. 
In experiments the tunnel conductances of the two junctions depend strongly
on the thickness of the oxide tunnel barriers.  We therefore
present numerical results for a variety of possible realizations of the
junction parameters.

We choose symmetric capacitances $C_{1}=C_{2}=C$ in our calculations. 
The important energy scale is then the Coulomb energy $E_{c}=e^{2}/2C$ and 
we scale the other relevant energies by this energy. The spin-dependent
junction conductances in the two junctions are described in units of a
typical junction conductance $G$, so that the electric current and the spin-current
are calculated in units of the typical current $Ge/2C$.

Let us first consider the DC transport properties.
In a calculation of the current through the system for a given source-drain
voltage $V$, we must first determine the non-equilibrium chemical potential
difference $\Delta \mu $ by the spin balance 
on the island (\ref{spmot}). The solution of this equation is given by the
intersection of the spin-current into or out of the island $I_{s}(\Delta \mu
)$ and the straight line $G_{s}2\Delta \mu /e$ which describes the spin-flip
relaxation in the island, where the spin-flip conductance $G_{s}$ was
defined in Eq.\ (\ref{scond}).
 
As we have pointed out in the previous chapter, the non-equilibrium
spin does not modify the Coulomb gap. However it does affect the
current when the source-drain voltage is larger than the Coulomb
gap. We show in Fig.\ \ref{f:cur} the current as a function of the
source-drain voltage $V$ for a system consisting of $G_{1\uparrow
}/G=0.3$, $G_{1\downarrow }/G=0.1$, $G_{2\uparrow }/G=3$,
$G_{2\downarrow }/G=6$, $C_{1}=C_{2}=C$, $q_{0}=0$ and the thermal
energy $k_{B}T=0.02E_{c}$. The upper curve shows the current when spin
relaxation is fast $G_{s}/G=1000$, and the lower curve shows the
current in the case of slow spin relaxation, $G_{s}/G=0$. The Coulomb
blockade threshold for low source-drain voltages is clearly seen and
the steplike structure for higher voltages called the Coulomb
staircase resulting from the discrete charging of the island
\cite{Grabert92}.  The Coulomb blockade and the Coulomb staircase are
smeared by the thermal fluctuations. Since the conductances of the
tunnel junctions are spin-dependent, the spin-flip relaxation on the
island is important for the current. The current increases with
increasing spin-flip conductance $G_{s}$ (decreasing spin-flip
relaxation time $\tau _{\mbox{sf}}$). In the case of fast spin-flip
relaxation ($G_s/G=1000$), the spin-accumulation is vanishing small,
and the steps occur at $eV=(2n+1)e^2/2C$, where $n=0,1,2,...$ . For
$G_s/G=0$, there is a spin- accumulation when $eV>e^2/2C$, so that the
relative energy difference associated with the tunneling of one
electron onto the island $E_i(V,q)-\sigma \Delta \mu/2$ is
spin-dependent. Hence, the spin-degenerate staircase at $eV=(2n+1)
e^2/2C$ for $n=1,2,3...$ splits into two peaks, where the splitting is
proportional to the spin-accumulation. This can be seen in the lower
curve in Fig.\ \ref{f:cur}, where the second staircase ($2n+1=3$)
occur at a voltage lower than $3 e^2/2C$ \cite{Imamura98}.

Another typical experiment on the single-electron transistor is to
measure the differential conductance in the linear source-drain
voltage regime as a function of the gate voltage $V_{g}$
($q_{0}=V_{g}C_{g}$). We show in Fig.\ \ref{f:cond} the influence of
the spin-flip relaxation in the island on the differential
conductance. The system is described by $G_{1\uparrow }/G=3.0$,
$G_{1\downarrow }/G=0.2$, $G_{2\uparrow }/G=0.1$, $G_{2\downarrow
}/G=2.0$, $C_{1}=C_{2}=C$ and the thermal energy is
$k_{B}T=0.1E_{c}$. The curves show the differential conductance as a
function of the gate voltage. The upper curve is calculated for fast
spin relaxation $G_{s}/G=500$, the mid curve is for intermediate spin
relaxation $G_{s}/G=0.5$ and the lower curve is for slow spin
relaxation $G_{s}/G=0.005$. The typical oscillatory dependence of the
differential conductance with respect to the gate voltage is seen. In
the same way as the current at high voltages increases with increasing
spin-flip relaxation in Fig.\ \ref{f:cur}, we see that the
differential conductance increases with increasing spin-flip
conductance (decreasing spin-flip relaxation time ). As usual the
peaks in the conductance as a function of the gate voltage increases
with decreasing temperature (not shown) \cite{Grabert92}.

The junction magnetoresistance is the relative difference in the
resistance on switching the directions of the magnetizations in the
leads from parallel to antiparallel (\ref{JMR}). Let us consider the
situation where the leads are ferromagnetic and the island is
non-magnetic, i.e. a F/N/F junction. The spin-accumulation causes a
non-zero magnetoresistance. In the parallel configuration, the
conductances are $G_{1\sigma }=G_{1}(1+\sigma P)/2$ and $G_{2\sigma
}=G_{2}(1+\sigma P)/2$, where $P$ is the polarization of the
ferromagnet. We show in Fig. \ref{f:tmr} the calculated junction
magnetoresistance for $G_{1}=G$, $G_{2}=3G$, $V_{g}=0$,
$k_{B}T=0.05E_{c}$ , $C_{1}=C_{2}=C$ and a polarization $P=0.3$ in the
limit of slow spin-flip relaxation $G_{s}=0$ (upper curve),
intermediate relaxation $G_{s}/G=1$ (mid curve) and fast spin-flip
relaxation $G_{s}=4G$ (lower curve). At low source-drain bias, we
observe the magnetoresistance oscillations already reported in \cite
{Barnas98:1058,Majumdar97:09206}. The amplitude of the oscillations
decreases with increasing source-drain voltage, thus decreasing
importance of the Coulomb charging
\cite{Barnas98:1058,Majumdar97:09206}. The period of the oscillations
for our system is close to $2E_{c}$. There is only a small distortion
of the shape of the magnetoresistance oscillations with increasing
spin-flip relaxation rate in the island. The oscillations in the TMR
as a function of the source-drain voltage can be understood as a
consequence of the spin-accumulation in the antiparallel
configuration. The spin-accumulation increases with increasing current
through the system. We have seen in Fig. \ref{f:cur} that the current
has a steplike behavior with a period close to $2E_c$ due to the
discrete charging of the island. Hence the spin-accumulation also
shows a steplike behavior (not shown) as a function of the
source-drain voltage. The magnetoresistance for the F/N/F FSET
increases with increasing spin-accumulation and hence shows
oscillations with a period $2E_C$.  The magnetoresistance is
noticeable even when the spin-flip conductance is of the same order as
the tunnel conductances in agreement with
Eq. (\ref{condition}). Disregarding the Coulomb charging energy, the
junction magnetoresistance is
\begin{equation}
\mbox{JMR}=P^{2}\frac{1-\gamma ^{2}}{1-P^{2}\gamma ^{2}+\alpha ^{2}},
\label{JMRFNF}
\end{equation}
where $\gamma =(G_{1}-G_{2})/(G_{1}+G_{2})$ is a measure of the asymmetry of
the junction conductances and $\alpha ^{2}=4G_{s}/(G_{1}+G_{2})$ determines
the reduction of the magnetoresistance due to the spin-flip relaxation. At a
high source-drain bias, the
numerical results agree well with Eq. (\ref{JMRFNF}), giving JMR$=6.9\%$ for 
$G_{s}/G=0$, JMR$=3.4\%$ for $G_{s}/G=1$ and JMR$=1.4\%$ for $G_{s}/G=4$.

So far we have shown how the spin accumulation influences the DC
transport properties.  Spin accumulation also causes novel features in
the AC and transient response of the system. Let us first fix the
source-drain voltage at a high bias until the system is stationary and
then lower the source-drain voltage. The transient current response in
this situation can be \emph{reversed} on time scales of the order of
the spin-flip relaxation time, which is an unambigous signature of a
non-equilibrium spin. In order to explicitly show this we solved the
time-dependent problem by numerically integrating Eq. (\ref {master})
and Eq. (\ref{spineq}). In the upper part of Fig. \ref{f:trcur} we
show the calculated time-dependent chemical potential difference when
the source-drain voltage is suddenly reduced at $t=0$.  The junctions
are described by $G_{s}/G=0.5$, $G_{1\uparrow }/G=0.05$,
$G_{1\downarrow }/G=1$, $G_{2\uparrow }/G=2$, $G_{2\downarrow
}/G=0.01$, $C_{1}=C_{2}=C$ and the thermal energy is
$k_{B}T=0.05E_{c}$.  We consider the high voltage case $V=8E_{c}$,
where the associated stationary electric current is $I=2.3Ge/2C$ and
lower the source-drain voltage to $V=2E_{c}$, where the associated
stationary electric current is positive, $I=0.40Ge/2C$.  We have used
$t_{\mbox{sf}}=20RC$, which appears to be a reasonable estimate for
junctions with a Coulomb charging energy of $10meV$ and a junction
conductance of $R/R_{K}=10$ giving a charge relaxation time of
$RC=0.5\cdot 10^{-12}s$. The chemical potential difference is seen to
decay on a time scale much larger than the charge relaxation
time. Finally we show in the lower part of Fig. \ref {f:trcur} the
current through the first junction (full line) and the second junction
(dotted line). It is clearly seen that the relaxation of the current
is slow on the time scale $RC$. For time scales up to the order of
$RC$, the currents through the first and the second junction differ
due to the depopulation of the charge island.  The Coulomb charging
effect shows up as the almost constant current on intermediate
time-scales.

In order to understand the dynamics it is useful to inspect the device
without the Coulomb charging effects, i.e. the capacitances $C_{1}$
and $C_{2}$ in the equivalent electric circuit in Fig.\
\ref{f:cir}. We set the voltage on the left lead to zero and apply a
time dependent potential $V(t)$ to the right lead. The complex
impedance $Z_{\mbox{spin}}(\omega )=V(\omega )/I(\omega )$ is
\begin{equation}
\frac{1}{Z_{\mbox{spin}}(\omega
)}=\frac{G_{1}G_{2}}{G_{1}+G_{2}}-\frac{G_{1\uparrow }G_{2\downarrow
}-G_{1\downarrow }G_{2\uparrow }}{(G_{1}+G_{2})} \frac{\Delta \mu
(\omega )}{eV(\omega )} \label{imp}
\end{equation}
where 
\begin{equation}
\frac{\Delta \mu (\omega )}{eV(\omega )}=\frac{1}{1+i\omega \tau
_{\mbox{spin}}}\frac{G_{1\uparrow }G_{2\downarrow }-G_{1\downarrow
}G_{2\uparrow }}{(G_{s}+G^{\prime })(G_{1}+G_{2})}.  \label{chem}
\end{equation}
Here the spin accumulation time is 
\begin{equation}
\tau _{\mbox{spin}}=\frac{C_{s}}{G_{s}+G^{\prime }},  \label{stime}
\end{equation}
where $1/G^{\prime }=1/(G_{1\uparrow }+G_{2\uparrow
})+1/(G_{1\downarrow }+G_{2\downarrow })$. From the relations
(\ref{imp}) and (\ref{chem}) we see why switching-off the source-drain
voltage ($V_{f}=0$) reverses the transient current as found in the
lower panel in Fig. \ref{f:trcur}. Without the Coulomb blockade this
transient decays on the time scale $\tau _{\mbox{spin}}$. In the limit
that the junction conductances are much smaller than the spin
conductance, the spin accumulation time (\ref{stime}) reduces to the
spin-flip relaxation time, $\tau_{\mbox{spin}}\approx \tau
_{\mbox{sf}}$. In the opposite limit where the junction conductances
are much larger than the spin conductance, the spin accumulation time
is $\tau _{\mbox{spin}}\sim C_{s}R$. The spin-flip capacitance is much
larger than the charge-capacitance $C$ in the regime where the
orthodox theory is valid ($\delta \ll E_{C}$) and thus the spin
accumulation time is much larger than the charge-relaxation time. The
spin accumulation time in Fig. \ref{f:trcur} agree well with the value
$t_{\mbox{spin}}=0.43t_{\mbox{sf}}$ as can be found from the
equivalent circuit neglecting the Coulomb blockade. The spin
accumulation time deviates from Eq. (\ref{stime}) if the initial or
final voltage is less than the Coulomb charging energy
\cite{Brataas98}. We show in Fig.\ \ref{f:trcur0} the average current
through the first and the second junction when the final voltage is
zero with the same system parameters as above (as used in Fig.\
\ref{f:trcur}). In this case the non-equilibrium spin accumulation
decays slower since the spins must relax through the spin-flip
conductance $G_{s}$ on the island and the transport through the
junctions is suppressed. The spin-accumulation time is then roughly
equal to the spin-flip relaxation time \cite {Brataas98}.  The
transient response to switching on the source-drain voltage is
similar. If the initial and final voltage are above the Coulomb
charging energy, the response time is determined by Eq. (\ref{stime})
otherwise the response time roughly equals the spin-flip relaxation
time.

The long time response of the system due to the spin dynamics can also
be observed in other AC transport experiments.  A fast single-electron
transistor has recently been realized\cite{Schoelkopf98:1238}. We
therefore study the influence of an AC source-drain voltage
$V(t)=V_0+V_1\cos\omega t$ on the DC current through the system,
$\bar{I}(\omega)=\omega/(2\pi)\int_{t_0}^{t_0+2\pi/\omega} dtI(t)$,
where $t_0$ is an arbitrary time constant. We show in Fig.\
\ref{f:freq} the relative change in the DC current
\begin{equation}
R(\omega)=\frac{\bar{I}(\omega=0)-\bar{I}(\omega)}{\bar{I}(\omega=0)}
\end{equation}
as a function of frequency for an applied voltage which fluctuates
around the Coulomb blockade threshold voltage, $V(t)=E_c(1+0.25\cos
\omega t)$. The temperature is $k_BT=0.05E_c$, $q_0=9$ and
$C_1=C_2$. The system is in the antiparallel configuration so that
there is a spin-accumulation on the island, $G_{1\uparrow}/G=1.5$,
$G_{1\downarrow}/G=0.5$,
$G_{2\uparrow}/G=0.5$,$G_{2\downarrow}/G=1.5$,$G_s/G=0.5$ and the
spin-flip relaxation time is $\tau_{\mbox{sf}}=20RC$. It is seen that
the DC current varies most strongly when $\omega \tau_{\mbox{sf}} \sim 1$,
which is a consequence of the spin-dynamics in the system with the
characteristic time-scale $\tau_{\mbox{sf}}$. A corresponding calculation
with the same parameters in the {\em parallel} configuration shows no
frequency dependence of $R(\omega)$ around $\omega \tau_{\mbox{sf}} \sim 1$.

\section{Conclusions}

\label{s:con}We have investigated the effect of non-equilibrium spins on
the transport properties of a ferromagnetic single-electron
transistor. The orthodox theory is generalized to include the spin
accumulation on the island. The spin accumulation is more important
for small islands with a large energy spacing.  The current and the
differential conductance increases with increasing spin-flip
relaxation rate. The magnetoresistance is enhanced due to the spin
accumulation. The non-equilibrium spins on the island can have a
drastic effect on the transient transport properties. We have shown
that on lowering the source-drain voltage from a high bias to a low
bias, a reversed current appears on time scales shorter than the
spin-flip relaxation time. The same slow response also appears in
other AC transport properties if one or more of the external
parameters are time-dependent.

\begin{acknowledgement}
This work is part of the research program for the ``Stichting voor
Fundamenteel Onderzoek der Materie'' (FOM), which is financially
supported by the ''Nederlandse Organisatie voor Wetenschappelijk
Onderzoek'' (NWO). We acknowledge benefits from the TMR Research
Network on ``Interface Magnetism'' under contract No. FMRX-CT96-0089
(DG12-MIHT) and the ``Monbusho International Scientific Research
Program on Transport and Magnetism of Microfabricated
Magnets''. G. E. W. B. would like to thank Seigo Tarucha and his group
members for their hospitality at the NTT Basic Research Laboratories
and Keiji Ono for a discussion.
\end{acknowledgement}

\begin{figure}[p]
\caption{The single electron transistor consisting of a (magnetic)
island coupled to two (magnetic) reservoirs by tunnel junctions.}
\label{f:set}
\end{figure}

\begin{figure}[tbp]
\caption{The non-equilibrium chemical potentials in the reservoirs and the
island at a finite source-drain voltage and a finite current.}
\label{f:non}
\end{figure}

\begin{figure}[tbp]
\caption{The equivalent circuit for the current-voltage response of the
system.}
\label{f:cir}
\end{figure}

\begin{figure}[tbp]
\caption{The electric current $I$ as a function of the source-drain voltage for
different spin relaxation conductances: $G_{s}/G=1000$ (upper line) 
 and $G_{s}/G=0$ (lower line). The system
parameters are $G_{1\uparrow }/G=0.3$, $G_{1\downarrow }/G=0.1$, $G_{2\uparrow }/G=3$, $G_{2\downarrow }=6$, $T=0.02E_{C}$, $q_{0}=0$, $C_{1}=C_{2}=1.0$.}
\label{f:cur}
\end{figure}

\begin{figure}[tbp]
\caption{The conductance $dI/dV$ as a function of the gate-voltage
$V_{g}$ for different spin-relaxation conductances: $G_{s}/G=500$
(upper line), $G_{s}/G=0.5$ (mid line) and $G_{s}/G=0.005$ (lower
line). The system parameters are $G_{1\uparrow }/G=3.0$,
$G_{1\downarrow }/G=0.2$, $G_{2\uparrow }/G=0.1$ , $G_{2\downarrow
}=2.0$, $T=0.1E_{C}$, $q_{0}=0$, $C_{1}=C_{2}=1.0$.}
\label{f:cond}
\end{figure}

\begin{figure}[tbp]
\caption{The magnetoresistance as a function of the source-drain
voltage in a F/N/F double junction system with P=0.3,
$G_{1\uparrow}+G_{1\downarrow}=G$, $G_{2\uparrow}+G_{2\downarrow}=3G$,
$V_g=0$, $C_1=C_2=1$ and $k_BT=1.0$.  The upper curve is for slow
spin-relaxation, $G_s/G=0$, the mid curve for intermediate spin-flip
relaxation $G_s/G=1$ and the lower curve for faster spin-relaxation,
$G_s/G=4$.}
\label{f:tmr}
\end{figure}

\begin{figure}[tbp]
\caption{The current as a function of time. The source-drain voltage
is switched from $V=8E_{c}$ to $V=2E_{c}$ at $t=0$. The system
parameters are $G_{s}/G=0.5$, $G_{1\uparrow }/G=0.05$, $G_{1\downarrow
}/G=1$, $G_{2\uparrow }/G=2$, $G_{2\downarrow }/G=0.01$,
$k_BT=0.05E_{C}$, $q_{0}=0$ and $C_{1}=C_{2}$.}
\label{f:trcur}
\end{figure}

\begin{figure}
\caption{The average current through the first and the second junction
as a function of time. The source-drain voltage is switched from
$V=8E_{c}$ to $V=0$ at $t=0$. The system parameters are $G_s/G=0.5$,
$G_{1\uparrow}/G=0.05$, $G_{1\downarrow}/G=1$, $G_{2\uparrow}/G=2$,
$G_{2\downarrow}/G=0.01$, $k_BT=0.05E_c$, $q_0=0$ and $C_1=C_2$}
\label{f:trcur0}
\end{figure}

\begin{figure}
\caption{The time-averaged (DC) current as a function of the frequency
of the applied source-drain voltage, $V(t)=E_C(1+0.25\cos\omega
t)$. The reservoirs are in the antiparallel configuration,
$G_{1\uparrow}/G=1.5$, $G_{1\downarrow}/G=0.5$,
$G_{2\uparrow}/G=0.5$,$G_{2\downarrow}/G=1.5$. The other parameters
are $k_BT=0.05E_c$, $q_0=9$ and $C_1=C_2$}
\label{f:freq}
\end{figure}


\begin{thebibliography}{99}
\bibitem{Grabert92} H. Grabert and M.~H. Devoret (eds.)
\textit{Single-charge tunneling} (Plenum Press, New York, 1992).

\bibitem{Kastner92:849}  M.~A. Kastner, Rev. Mod. Phys. \textbf{64}, (1992)
849.

\bibitem{Gijs97:285} M.~A.~M. Gijs and G.~E.~W. Bauer,
Adv. Phys. \textbf{46}, (1997) 285.

\bibitem{Levy94:367}  P.~M. Levy, Solid St. Physics \textbf{47}, (1994) 367.

\bibitem{Ono96:3449}  K. Ono, H. Shimada, S. Kobayashi, and Y. Ootuka, J.
Phys. Soc. Jpn \textbf{65}, (1996) 3449.

\bibitem{Ono97:1261}  K. Ono, H. Shimada, and Y. Ootuka, J. Phys. Soc. Jpn 
\textbf{66}, (1997) 1261.

\bibitem{Takahashi98:1758}  S. Takahashi and S. Maekawa, Phys. Rev. Lett. 
\textbf{80,} (1998) 1758.

\bibitem{Shimada97:11075}  H. Shimada, K. Ono, and Y. Ootuka,
cond-mat/9711075 (1997); K. Ono, H. Shimada and Y. Ootuka, cond-mat/980370 (1998). 

\bibitem{Sankar97:5512}  S. Sankar, B. Dieny, and A.~E. Berkowitz, J. Appl.
Phys. \textbf{81}, (1997) 5512.

\bibitem{Schelp97:R5747} L.~F. Schelp \textit{et~al.}, Phys. Rev. B
\textbf{56}, (1997) R5747.

\bibitem{Barnas98:1058}  J. Barnas and A. Fert, Phys. Rev. Lett. \textbf{80,}
(1998) 1058.

\bibitem{Majumdar97:09206}  K. Majumdar and S. Hershfield, cond-mat/9709206
(1997).

\bibitem{Johnson85:1790}  M. Johnson and R. H. Silsbee, Phys. Rev. Lett. 
\textbf{55}, (1985) 1790.

\bibitem{Tedrow94:174} P. M. Tedrow and R. Meservey,
Phys. Rep. \textbf{238} , (1994) 174.

\bibitem{Johnson93:2142}  M. Johnson, Phys. Rev. Lett. \textbf{70}, (1993)
2142.

\bibitem{Abrikosov62:1088} A. A. Abrikosov and L. P. Gor'kov, Zh. Eksp. Teor.
Fiz. \textbf{42}, (1962) 1088 [Sov. Phys. JETP \textbf{15}, 1962 (752)].

\bibitem{Meservey78:805} R. Meservey and P. M. Tedrow, Phys. Rev. Lett. 
\textbf{41}, (1978) 805.

\bibitem{Tans98:49} S. J. Tans, M. H. Devoret, R. J. A. Groeneveld and
C. Dekker, Nature \textbf{393}, (1998) 49.

\bibitem{Matsukara98:R2037}  F. Matsukara and H. Ohno and A. Shen and Y.
Sugawara, Phys. Rev. B \textbf{57}, (1998) R2037.

\bibitem{Brataas98}  A. Brataas, Yu. V. Nazarov, J. Inoue and G. E. W.
Bauer, unpublished.


\bibitem{Julliere75:225} M. Julliere, Phys. Lett. \textbf{54A}, (1975) 225. 


\bibitem{Grabert91:143}  H. Grabert \textit{et~al.}, Z. Phys. B \textbf{84},
(1991) 143.

\bibitem{Imamura98} A. B. is grateful to H. Imamura for a remark.

\bibitem{Schoelkopf98:1238} R. J. Schoelkopf, P. Wahlgren,
A. A. Kozhevnikov, P. Delsing and D. E. Prober, Science \textbf{280},
(1998) 1238.

\end{thebibliography}
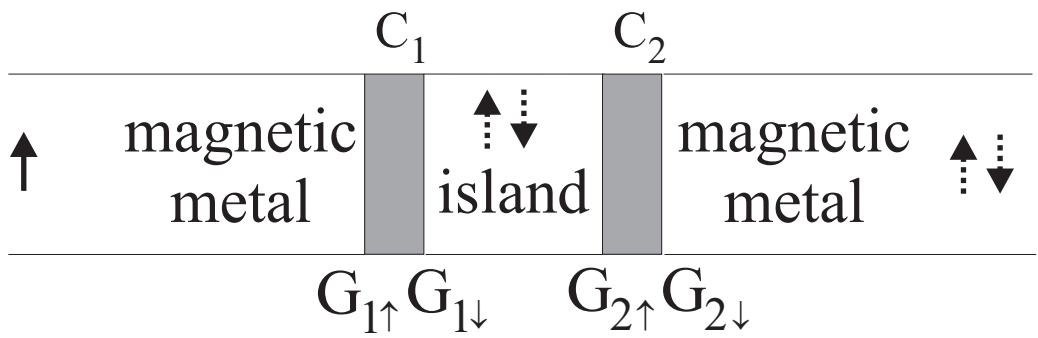
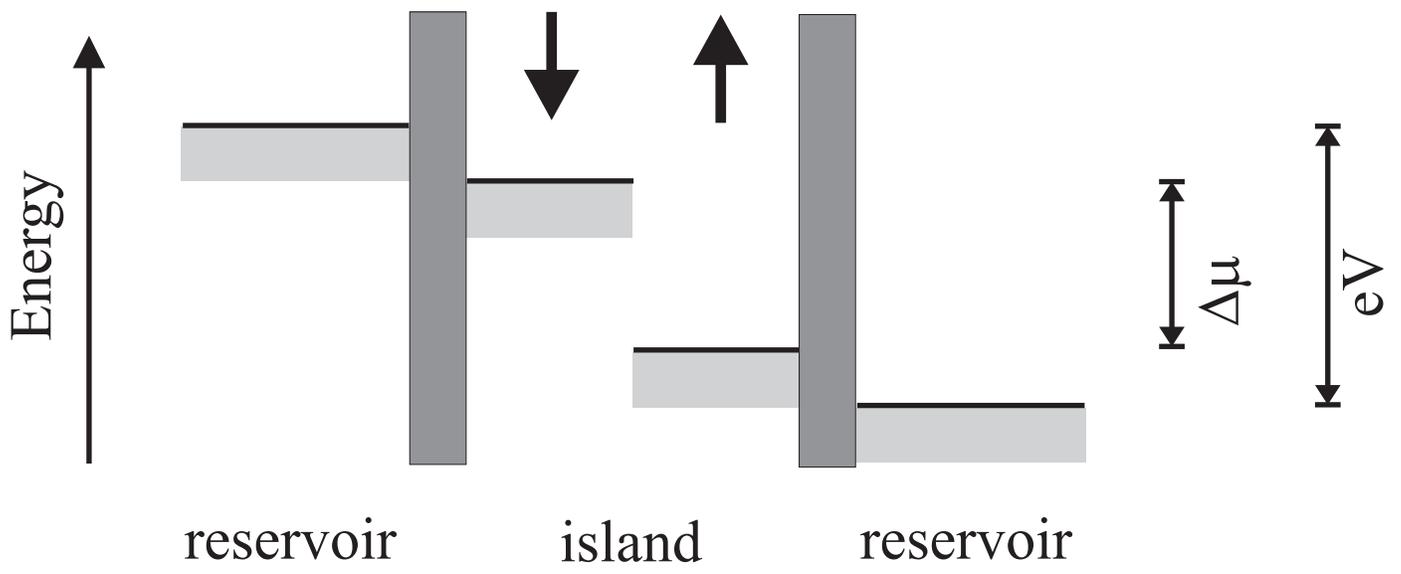
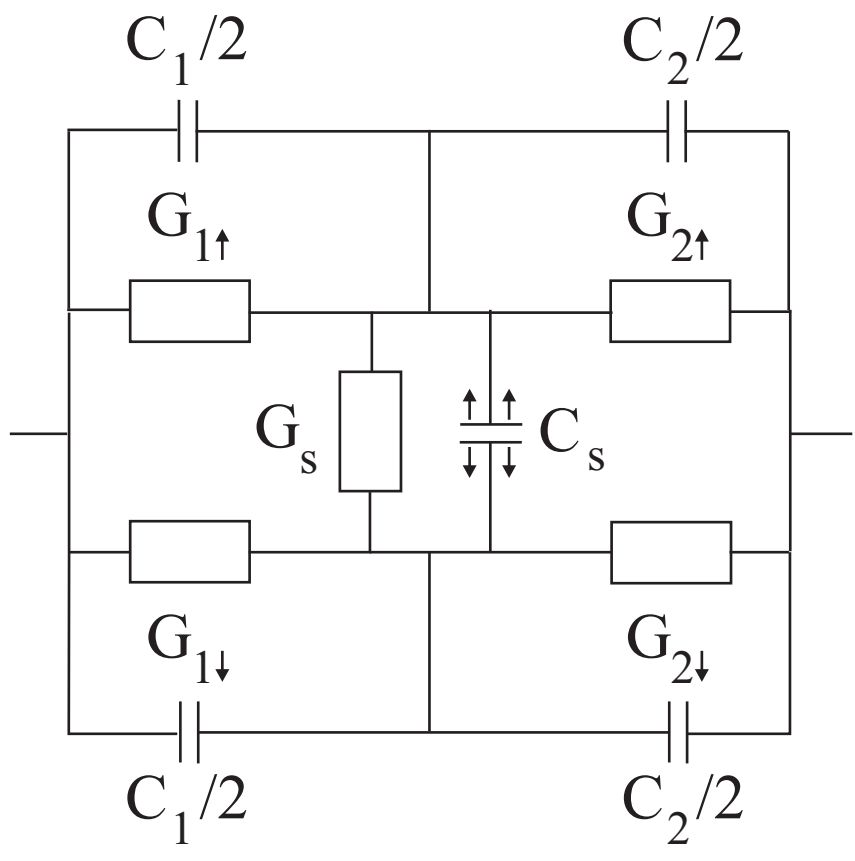
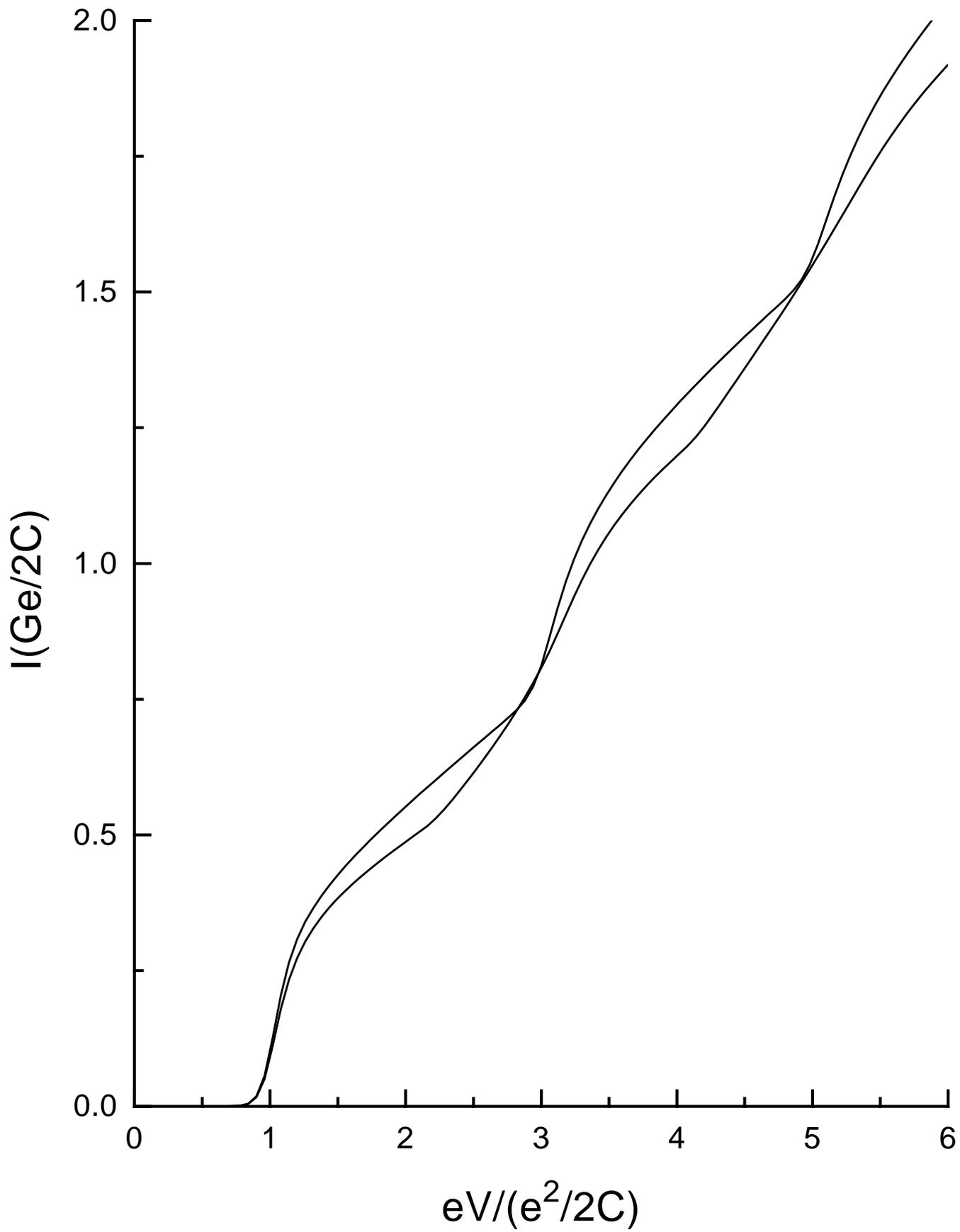
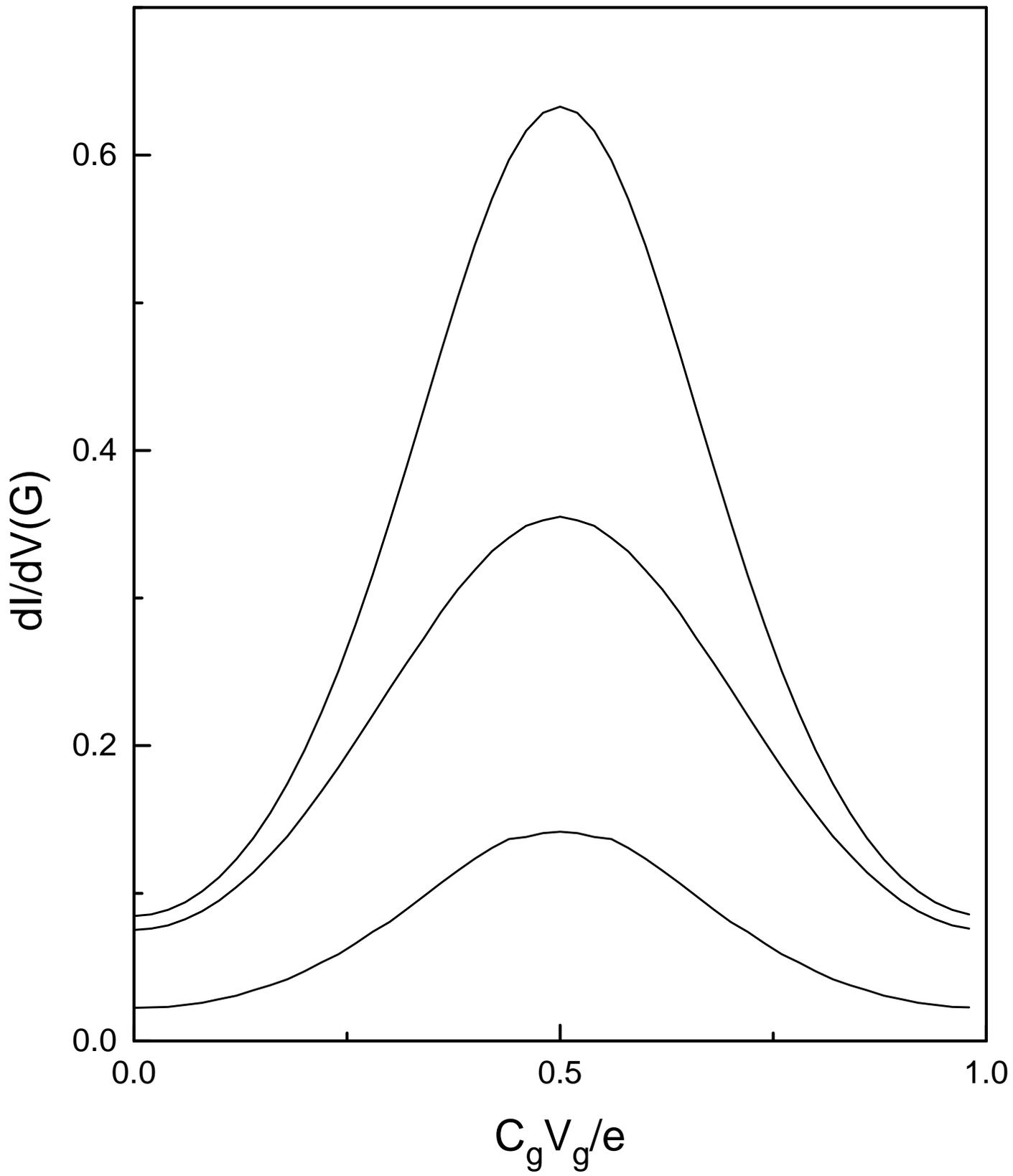
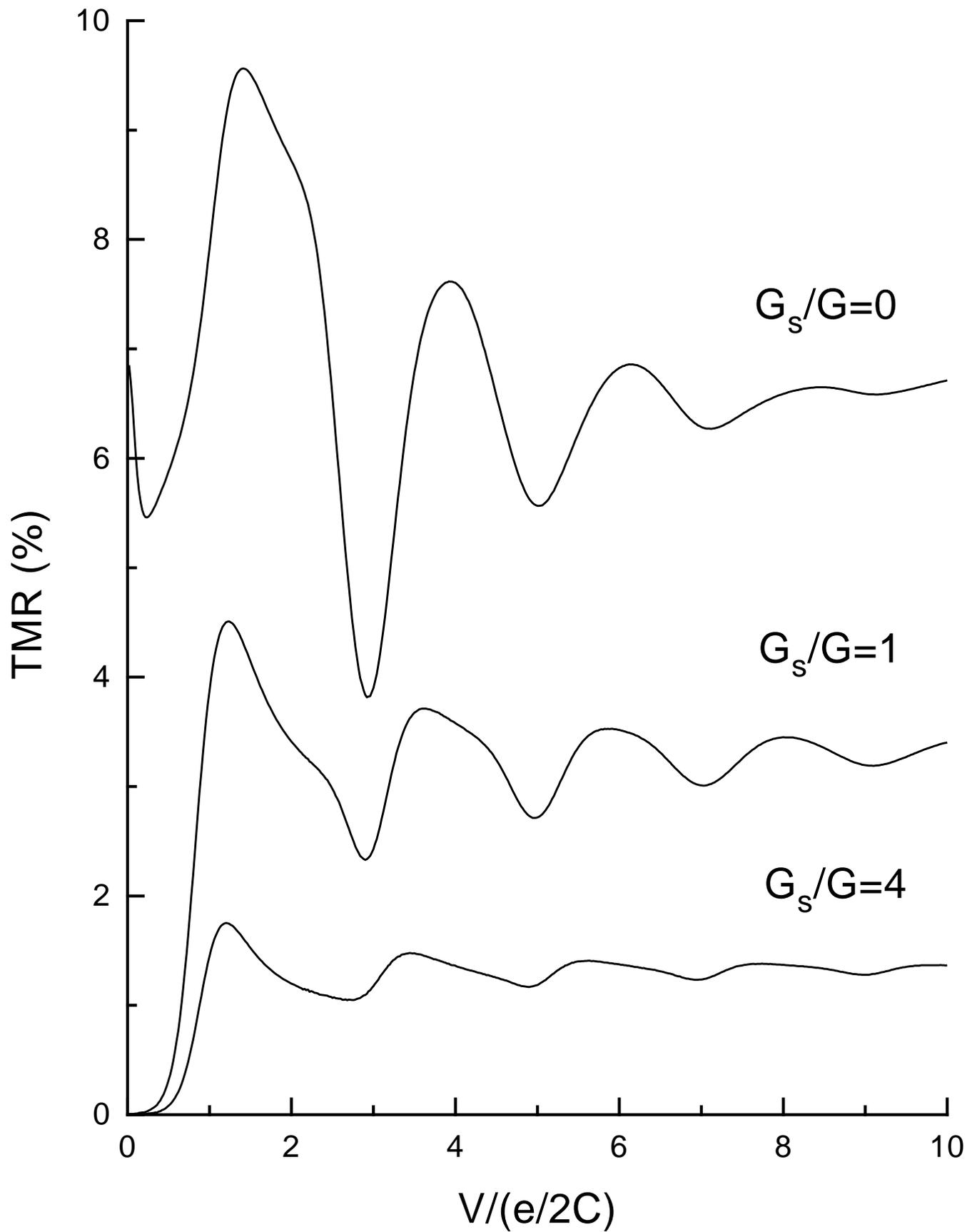
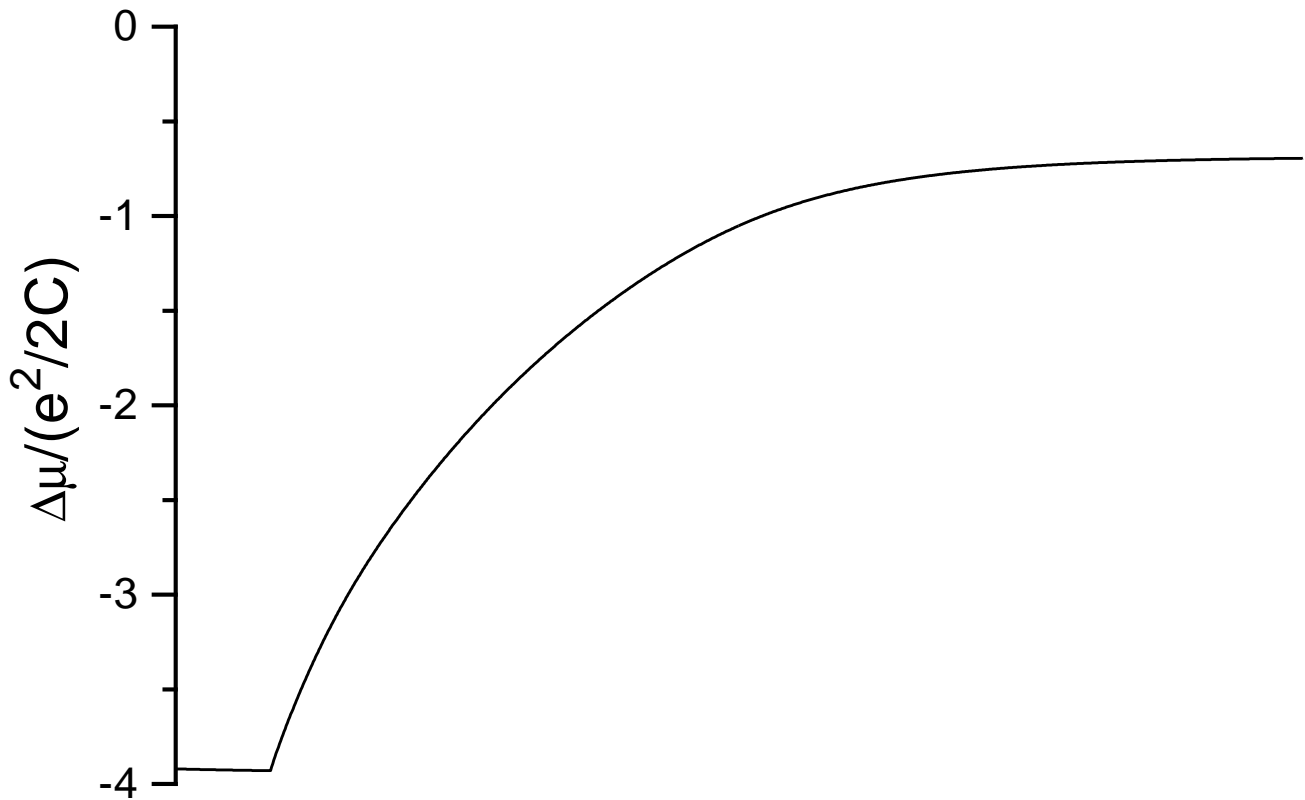
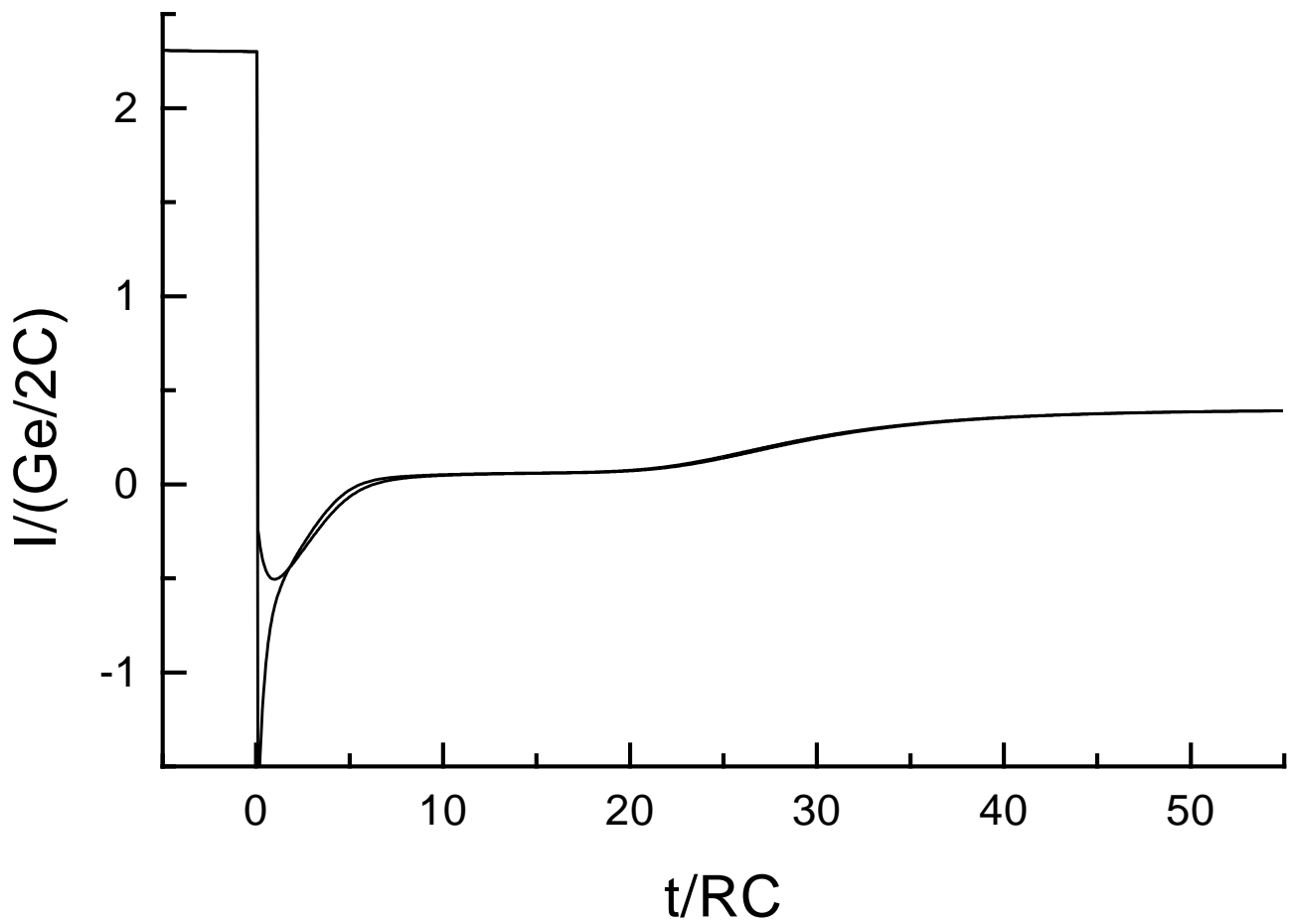
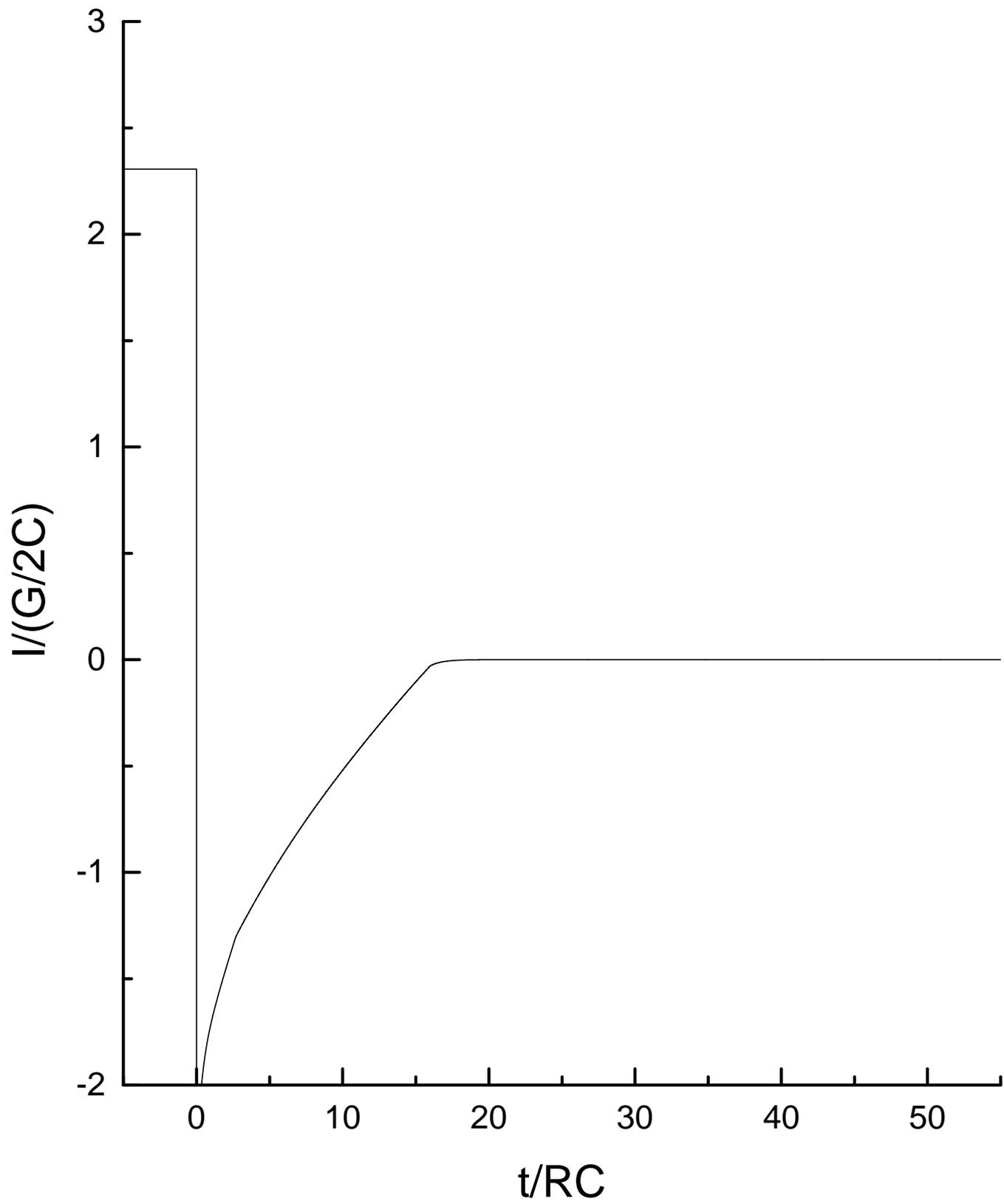
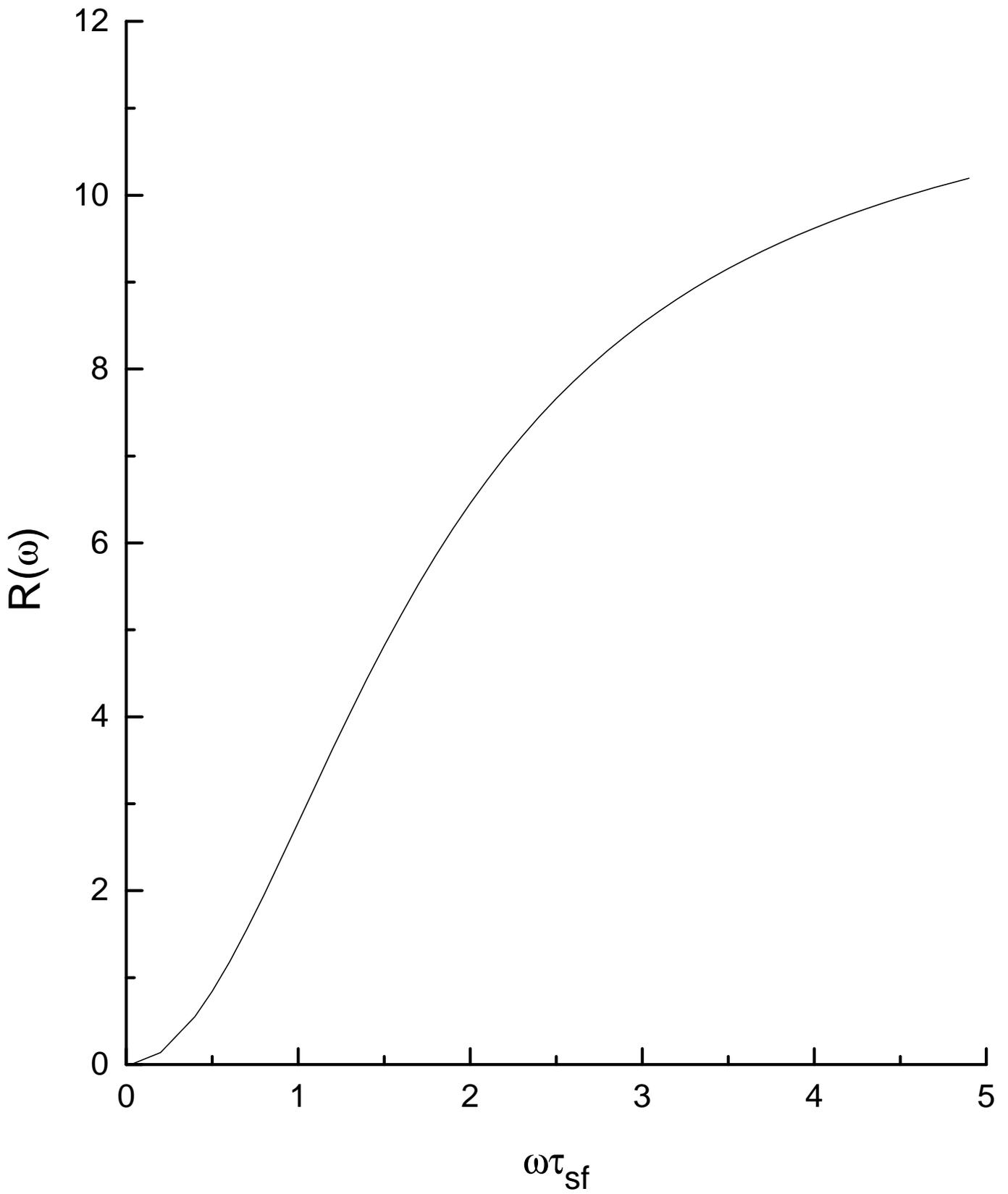
\end{document}